\def\kms{\mbox{km sec$^{-1}$}}
\def\kpc{\mbox{kpc}}
\def\msun{\mbox{M$_\odot$}}
\def\gcm{\mbox{g cm$^{-3}$}}
\def\sige{\mbox{$\sigma_8$}}
\def\omeone{\mbox{$\Omega_0 = 1.0$}}
\def\ometwo{\mbox{$\Omega_0 = 0.2$}}
\def\ome{\mbox{$\Omega_0$}}
\def\lam{$\lambda_0$}
\def\eps{\mbox{$\epsilon$}}
\def\epsrho{\mbox{$\epsilon_{\rho \rho}$}}
\def\epshh{\mbox{$\epsilon_{h h}$}}
\def\mpc{\mbox{Mpc}}
\def\sigrho{\mbox{$\sigma_{1, \rho \rho}$}}
\def\sighh{\mbox{$\sigma_{1, h h}$}}
\def\sigg{\mbox{$\sigma_{g g}$}}
\def\xirho{\mbox{$\xi_{\rho \rho}$}}
\def\xihh{\mbox{$\xi_{h h}$}}
\def\xigg{\mbox{$\xi_{g g}$}}
\def\xihha{\mbox{$\overline\xi_{hh}$}}

\def\sigei{\mbox{$\sigma_8(t_i)$}}

\def\mathnew{\mathsurround=0pt}
\def\simov#1#2{\lower .5pt\vbox{\baselineskip0pt
    \lineskip-.5pt\ialign{$\mathnew#1\hfil##\hfil$\crcr#2\crcr\sim\crcr}}}  
\def\simgreat{\mathrel{\mathpalette\simov >}}
\def\simless{\mathrel{\mathpalette\simov <}}
\def\'#1{\ifx#1i{\accent"13\i}\else{\accent"13#1}\fi}
\def\eg{e.g.}
\def\et{et~al.}
\def\ie{i.e.}

\documentstyle[11pt,aaspp4]{article}

\begin{document}

\title{The \ome\ dependence of the evolution of $\xi(r)$}

\author{P. Col\'in\altaffilmark{1}}
\affil{Instituto de Astronom\'ia, Universidad Nacional Aut\'onoma
de Mexico, Mexico}

\author{R. G. Carlberg}
\affil{Department of Astronomy, University of Toronto, 
       Toronto ON, M5S~3H8 Canada}

\and 

\author{H. M. P. Couchman}
\affil{Department of Astronomy, University of Western Ontario,
        London, ON, N6A~3K7 Canada}

\altaffiltext{1}{This work was done when he was a postdoctoral fellow
of the Department of Astronomy, University of Toronto}

\begin{abstract}
The evolution of the two-point correlation
function, $\xi(r,z)$, and the pairwise velocity dispersion,
$\sigma(r,z)$, for both the matter, \xirho, and halo population,
\xihh, in three different cosmological models: (\ome,\lam)=(1,0),
(0.2,0) and (0.2,0.8) are described. If the evolution of $\xi$ is 
parameterized by $\xi(r,z)=(1+z)^{-(3+\eps)}\xi(r,0)$, where
$\xi(r,0)=(r/r_0)^{-\gamma}$, then $\epsrho$ ranges from
$1.04 \pm 0.09$ for (1,0) and $0.18 \pm 0.12$ for (0.2,0), 
as measured  by the
the evolution of \xirho\ at
1 Mpc (from $z \sim 5$ to the present epoch). 
For halos, \eps\ depends indeed on their mean overdensity.
Halos with a mean overdensity of about 2000 were used to
compute the halo two-point correlation function, \xihh,
tested with two different group finding algorithms: the {\it 
friends of friends} and the spherical overdensity algorithm. It is
certainly believed that the rate of growth of this \xihh\ will give a good 
estimate of the evolution of the galaxy two-point correlation function,
at least from  $z \sim 1$ to the present epoch.
The values we get for \epshh\ range from 1.54 for (1,0) to  
-0.36 for (0.2,0), as measured by the evolution of
\xihh\ from $z \sim 1.0$ to the present epoch. These
values could be used to constrain the cosmological scenario.

The evolution of the pairwise velocity dispersion for the mass and
halo distribution is measured and compared with the evolution
predicted by the Cosmic Virial Theorem (CVT). According to the CVT,
$\sigma(r,z)^2 \sim G Q \rho(z) r^2 \xi(r,z)$ or $\sigma \propto
(1+z)^{-\eps/2}$. The values of $\eps$ measured from our simulated
velocities differ from those given by the evolution of $\xi$ and the CVT,
keeping $\gamma$ and $Q$ constant:
$\eps = 1.78 \pm 0.13$ for (1,0) or $\eps = 1.40 \pm 0.28$
for (0.2,0).

\end{abstract}
\section{Introduction}

The large scale structure of the Universe that we see today 
is believed to have developed from the growth of small
perturbations in the matter density driven by gravitational instability. 
The
evolution of the clustering of the mass density field depends on the
initial conditions via the density power spectrum and the
mean density of the universe and is therefore a powerful constraint
on theories of structure formation. 
The evolution of the galaxy clustering, however, need not
necessarily follow that of the collisionless component of the
mass density field. Galaxies have been subject to external
phenomena such as tidal interactions, satellite accretion, mergers,
etc. or internal phenomena such as galactic winds, 
that it would be unlikely to see
the galaxy clustering evolution being the same as the clustering
evolution of the dark matter.

Two straightforward statistical tools which describe the
clustering properties of galaxies, positions and velocities, are the 
two-point correlation function, $\xi(x,t)$, and the pairwise velocity
dispersion, $\sigma(x,t)$, hereafter comoving coordinates are denoted
by $x$ while proper coordinates are denoted by $r$. 
In a flat universe where initial conditions
were generated by a power-law spectrum with spectral index
$n$ the two-point correlation
function should scale as 
\begin{equation}
\xi(x,t)= \xi(s),
\end{equation}
where $s= x/t^\alpha$ and $\alpha = 4/[3(3+n)]$ (\cite{peeb}). 
Furthermore, it has been shown that 
even in the case where 
the hypothesis of scaling is broken, for instance, by a
scale-dependent power spectrum, relation (1) has proved to be a very
good approximation (\eg\ \cite{ham}; \cite{paddy}). 

In the regime where the density perturbations grow linearly
\begin{equation}
\xi(r,t) = b(t)^2 \xi(r,t_i),
\end{equation}
where $b(t)$ is the growing
mode of the density perturbations and $\xi(r,t_i)$
is the initial correlation function ($b(t_i) = 1$). 
In the particular case of \omeone, where $b$ is just
the expansion factor of the universe, $a = (1+z)^{-1}$,
and $P(k) \propto k^n$ : $\xi(x,z) \propto (1+z)^{-2} x^{-(n+3)}$
or $\xi(r,z) \propto (1+z)^{-(n+5)} r^{-(n+3)}$. 
On the other hand, 
in the highly non-linear regime where the 
hypothesis of stable clustering is supposed to work, $\xi(r,z)=
\xi(r,0) (1+z)^{-3}$. A convenient form of
parameterizing the evolution of $\xi\ $ is 
\begin{equation}
\xi(r,z) = (1+z)^{-(\eps+3)} \xi(r,0), 
\end{equation}
as it removes the universal expansion (\cite{gpeeb}).
If the hypothesis of
stable clustering is satisfied $\eps = 0$ while
$\eps = 2+n$ in the linear regime (\omeone\ and 
$P(k) \propto k^n$).

The present observed galaxy two-point 
correlation function  $\xi(r,0)$ is to a good approximation a
power-law $\xi_0 = (r/r_0)^{-\gamma}$. \cite{dp} from the CFA survey 
find $\gamma = 1.77 \pm 0.04$ and
$r_0 = 5.4 \pm 0.3 h^{-1} 
\mpc$. \cite{lmep} from the Stromlo--APM survey find 
$\gamma = 1.71 \pm 0.05$ and $r_0 = 5.1 \pm 0.2 h^{-1} \mpc$. Values
for $\gamma$ consistent with those found locally have been measured at
moderate redshifts (Shepherd \et\ 1996; \cite{cfrs}).  

It is seen from the studies by \cite{shep_cnoc} and Le F\`{e}vre \et\ (1996)
that the correlation length, $r_0$, has evolved a great deal.
Le F\`{e}vre \et\ find that $r_0$ has decreased by a
factor of 10, assuming $\ome = 1$, from the present epoch to $z \sim 0.6$.
An $\eps_{g g} \sim 1 \pm 1$ is derived from
these two studies.

There exists an extensive literature both observational
and theoretical that at least mention the name of $\xi$.
Here we are interested in those works that deal directly
with the time dependency of the two-point correlation function
for both the density field and the halo population. In this sense,
the paper by Davis \et\ (1985) is pioneer. They studied, among 
other things, the evolution of \xirho\ and `\xigg'\ in different
cosmological models, their numerical simulations consisted
of $32^3$ particles in a $64^3$ grid. They found no way of reproducing
simultaneously the observed amplitude and slope of the
correlation function with \xirho\ in their (\ome,\lam)=(1.0,0.0) 
model (a model hereafter
is represented by a pair of coordinates, where the first coordinate is 
\ome\ and the second coordinate \lam).
A better match was obtained with their low-density models (flat
and open). A biased galaxy formation scenario was invoked to
save the (1.0,0.0) model and a
\xigg\ was computed. Their \xigg\ was always above of their corresponding 
\xirho. Much of the clustering of these `halos' was due to 
the pattern imposed by the initial conditions. This work
was not intended to study the evolution of $\xi$ although
it certainly showed it. On the other hand, the evolution
they found for \xigg\ is certainly a rough approximation
because we know halos do not necessarly arise from high
peaks and high-peak particles do not necessarly end up in
halos. The evolution of $\xi$ has also been showed and
studied by other authors (\eg\ \cite{carl91}; \cite{bv1};
\cite{bv2}). Carlberg used fof
to compute the evolution of \xihh, and found that 
while the correlation length of the density field 
continues to grow 
in comoving coordinates, \xihh\ did not change. 
Brainerd \& Villumsen (1994), hereafter BV,
with the analysis of simulations
of $128^3$ particles, in a standard CDM scenario, 
were deeper in redshift
(they started their analysis at $z=5$ as opposed to $z=2.15$ by
Carlberg) and found a non-monotonic growth of \xihh.
Halos found by fof initially produce a biased \xihh\ (they follow
closely the large-scale pattern of filaments and sheets just
as particles located in high-density peaks did) and then
decreases because of mergers. The shape of \xihh\ will be mainly set
by four competing phenomena: (1) mergers, (2) formation,
(3) dynamics, and (4) disruption. By the present epoch, 
because mergers dominates the scene,
one would expect to have a \xihh\ that lies below \xirho, how high
will this bias be? is a question whose answer will depend on 
how much are these halos affected
by merging and disruption (this will be of course an environmental
effect).

In neither of the studies mentioned above was a value for $\eps$ computed.
Recently, two authors, at least, have computed the rate of evolution 
of \xirho\ using equation (3). Jain (1996) computed \eps\ as a function
of $r/r_0(a)$ for a standard CDM scenario. 
He founds values for \eps\ that range from $\sim -0.4$ to
$\sim 2.0$. The rate of growth could be much faster that those 
commonly cited numbers 0 or $\sim 1$. Peacock (1996) found,
on the other hand, that the rate of
growth in an open universe could solve the apparent contradiction
that: (1) the slope of \xigg\ seems not to change at all up to
$z \simeq 1$ and (2) the rate of evolution is relatively rapid
with an \eps\ value close to 1. 

In dissipationless N-body simulations a great effort has been made to
find collapsed objects which subsequently  could be asociated
with real galaxy halos ( \cite{defw}; \cite{cc}; \cite{bgelb};
\cite{wqsz}; \cite{lc}; \cite{sde}; \cite{van} ; \cite{knp}). In particular,
a still often used group finding algorithm is {\it friends of friends},
fof (\cite{defw}). This algorithm
find groups of particles, halos, that are connected more closely
than a specified link length, $l$ (\ie\ particles that are in an 
overdensity region
in excess of $\delta_{min} \sim [\frac{4\pi}{3}(l/2)^3]^{-1} \sim 2/l^3$).
It is well known that it suffers from the defect of 
joining, once in a while,
two halos that are physically distinct. To avoid this problem
others group finding algorithms were developed. One of these is
the spherical overdensity algorithm described by Lacey \&
Cole (1994). Although the identification of these collapsed objects
with galaxy halos seems at first not a bad approximation, it was soon
realized that they suffered from the defect of {\it overmerging}
(\eg\ \cite{fwde}). Most of these halos identified in an 
early epoch will be destroyed
by the time they reach the present epoch and this is a function
of the mean overdensity of the halo and mass resolution of the
simulation (halos with the same mass are more tight in
simulations with better mass resolution). 
Some very nice algorithms have been invented to solve this
problem and here we cite again the paper by Summers \et\ where a complete
discussion about this problem can be found (see also \cite{van}). 
However, they are very model dependent.
In this paper we try to avoid the problem of overmerging by
considering only halos whose mean overdensity
is rather high ($\sim 2000$). We still expect the clustering
of halos be less than the clustering of galaxies at recent epochs,
specially at small scales where very massive
halos composed of only one particle should contain many 
galaxies.
Therefore, the analogy 
of halos with real galaxy halos should not be taken beyond
its scope. The statistics of halos or, other galaxy
tracers, has been extensively studied at the present epoch and compared
with observations (\eg\ \cite{wfde}; \cite{cc}; \cite{gelbb}).   

An estimate of the galaxy pairwise velocity dispersion is given by
the Cosmic Virial Theorem (CVT, \cite{peeb})
\begin{equation}\sigma(r,z)^2 = \frac{3 J(\gamma) H(z)^2 \ome(z)Q r_0(z)^\gamma
  r^{2-\gamma}}{4 (\gamma-1)(2-\gamma)(4-\gamma)}
\end{equation}
where $Q$ is the three-point parameter ($J(1.7)=4.14$). 
By assuming that $Q$ and $\gamma$ do not vary with $z$ and
that $\xi \propto (1+z)^{-(3+\eps)}$, the behavior of $\sigma(z)$
is required to be: $\sigma \propto (1+z)^{-\eps / 2}$. An estimate of
$\sigg (h r = 1 \mpc)$ from the CFA survey by \cite{dp}
gives $300 \pm 40 \kms$, which in turn produces a value for $\ome =
\rho_0 / \rho_c$ ($\rho_c =  3 H_0^2 / 8 \pi G = 1.879 h^2 \times 
10^{-29} \gcm$, $h$ is the Hubble constant in units of 100 \kms
\mpc$^{-1}$); however, the value of $\sigg$ can be as high as $\sim 1000\ \kms$
if Coma is included (\cite{mjb}). Because
galaxies may not dynamically represent the background mass density
field, the accuracy of the CVT
as an estimator of \ome\ depends on how well galaxies follow the
background dark matter. 

This paper is focused to three main goals: (1) do a more systematic
analysis of the evolution of $\xi$ by computing it for both the
density field and a halo population (tested with two group finding 
algorithms), with a high mean overdensity, in three cosmologicals
models: (1.0,0), (0.2,0), and (0.2,0.8). (2) Include 
the evolution of the pairwise velocity dispersion. (3) Put results
in a convenient manner so as an observer can use them as they are
in the paper.
The outline of the paper is as follows. In \S2 the characteristics
of the simulations are described. In \S3 the evolution of 
\xirho\ is discussed and values for the
parameter $\eps$ are given. In \S4 the evolution of
\xihh\ is discussed. In \S5 the evolution of the
first and second moment of the mass and halo velocity field is
presented. And finally in \S6 a summary is presented.

\section{Numerical Techniques}

The simulations were performed using 
the adaptive particle-particle, particle-mesh,
AP$^3$M, N-body code of \cite{couch}. Each
simulation consisted of $128^3$ particles in a $128^3$ grid.

The initial conditions are generated using the Zeldovich
approximation as described by Efstathiou et al. (1985). The CDM input 
density power spectrum, $P(k)$, is the fit given by \cite{ebw}
\begin{equation}
P(k) = \frac{ A k}{\left(1 + [ak + (bk)^{3/2} + (ck)^2]^\nu \right)^{2/\nu}},
\end{equation}
where $A$ is a normalization constant, $a= 6.4/\Gamma$, $b=3.0/\Gamma$,
$c = 1.7/\Gamma$, and $\Gamma = \ome h$. The value of $h$ was set to
1 for (1.0,0.0) and (0.2,0.0) while for (0.2,0.8) it was set to 0.7. 
Therefore $\Gamma$ ranges from 0.14 for (0.2,0.8) to 1 for (1.0,0.0).
The power spectrum is normalized to $\sige = 1$ in
(1.0,0.0) and (0.2,0.0), and to COBE in (0.2,0.8).
\sige\ is the linearly predicted {\it rms}
mass fluctuation in an $8  h^{-1}$ Mpc sphere at the present time. 
One is left with freedom  of choosing any initial value for \sige\, subject 
only to the constraint that Zeldovich approximation is applicable.
We ran several experiments with $64^3$ particles. These 
indicated that by starting a simulation with a rather high initial
\sige, \sigei\ (a value of 0.3 would be considered
high), one might underestimate the present pairwise 
velocity dispersion by up to 30\% to 40\%, depending on the value 
of \ome. To make sure that
our preferred value of 0.1 for \sigei\ did not suffer from this effect,
we ran two simulations with two different values for \sigei: 0.025 and
0.1 for the \omeone\ model. 
Negligible differences were found in the two-point correlation function
and the pairwise velocity dispersion at $z=0$ between both simulations.  
All runs are initialized to $a(t_i = 1)= 1$, where $t_i$ is the initial
time in grid units and $a$ is the expansion factor. 
The force law corresponds to that between two finite
spherically symetric density clouds with shape given by
$ \rho(r) = (48/\pi \eta^4)(\eta/2 -r)$ for $r < \eta/2$, 
where $\eta$ is the smoothing length. The value of $\eta$
is constant in proper coordinates, with a value at 
the present epoch, $\eta_0 = 50\ \kpc$.
The number of timesteps [1,000 for (1.0,0.0), 1,650 for (0.2,0.0), and 1,000
for (0.2,0.8)] was chosen sufficiently high to satisfy 
the stability criteria of the numerical integration
(see, for example, \cite{edfw}) at all times. In fact, 
because the softening
parameter $\eta$ is kept constant in physical units in our simulations, 
we require only that $dt \ll min(\sqrt{\frac{6\pi}{4}} (\eta   a)^{3/2}, 
3t/2)$ (expression valid for $\ome = 1$), where $\eta$ and $a$ 
are in grid units.
The (0.2,0.0) and (0.2,0.8) both started at $\sigei = 0.1$ 
and had an expansion factor of 27.4 and 14.1, respectively.

\section{Evolution of $\xirho(r)$}

The two-point correlation function was measured by a
direct summation of pairs in bins as given by the formula
\begin{equation}
\xi(r)=(N_p/\bar n N_cdV) - 1,
\end{equation}
where $N_p$ is the number of pairs between $r$ and $r + dr$, $dV$ is
the volume of this spherical shell, $N_c$ is the number of particles
taken as centers, and $\bar n$ is the mean number density of
particles. The correlation function is neither well determined to
distances greater than about one
tenth of the comoving size of the box nor to distances smaller than
twice the force resolution.

\subsection{Evolution of $\xirho$ and the Parameter $\eps$}

The evolution of $\xirho$ in physical coordinates is plotted in 
Figure~\ref{fig:xirho1} for (1.0,0.0), in Figure~\ref{fig:xirho2} for 
(0.2,0.0), and in Figure~\ref{fig:xirho3} for (0.2,0.8).
The straight line is a power-law of exponent $-1.8$ and epochs are
labeled by different symbols: present epoch (+ symbol), $z \sim 0.5$ 
(* symbol), $z \sim 1.0$ (open circle), $z \sim 1.5$ (x symbol), 
$z \sim 2.8$ (open square), and $z \sim 5.0$ (open triangle). 
The symbol $\sim$ was used because the correct redshift depends 
on the values of \ome\ and \lam\ (see Table~\ref{tab:xirho} below). These
epochs have nothing of special.

If one parameterizes the evolution of $\xirho$ by 
$\xirho(x,z) = [x_0(z)/x]^{\gamma(z)}$ ($x$ denotes comoving coordinates), 
where $x_0(z)$
is the correlation length as a function of redshift, 
a value for $\eps$ can be derived by assuming that $r_0(z) \propto
(1+z)^{-(3+\epsilon)}$. The correlation length and the slope, $\gamma$,
is then computed by fitting a straight line to 
$\log \xi$ {\it vs.} $\log x$ in the
range $[\eta,x_0]$. In Table~\ref{tab:xirho} 
are shown our results. In the first and 
second columns are shown the redshift of the model and the correlation 
length in comoving units (as defined by $\xirho(x_0) = 1$). 
In the third column is shown $r_0$ in physical units. The correlation 
length computed by using the fit is shown in the fourth column along
with its 1$\sigma$ error, column five. 
The value of $\gamma$ and its uncertainty
are located in the sixth and seventh columns, respectively. The
values of \xirho\ at 1 \mpc\ and 0.2 \mpc\ are shown in columns
eight and nine, respectively.
From this table we see that the correlation length measured by 
assuming that $\xirho$ is a power-law is higher than the
one defined by $\xirho(r_0)=1$ (if \xirho\ were a perfect 
power-law both would coincide), \ie\ there appears to be 
more than one slope (\omeone), one from $1 < \xirho \simless 100$ 
and the second from $ 100 \simless \xirho < \xi_{\eta}$, 
where $\xi_{\eta}$ denotes
\xirho\ at the softening. (This latter regime likely corresponds
to stable clustering.)
This effect decreases as we go to earlier
epochs and this agrees with the idea that the highly non-linear 
regime bends \xirho. The highly non-linear effect decreases
as we go to higher redshifts for two reasons: (1) less intrinsic
clustering and (2) our scheme that fixes the resolution in physical
coordinates. 

The $\eps$ parameter is computed by simply fitting the $\log r_0\ vs.\
\log (1+z)$ ($r_0$ as measured by the fit), 
from where one can extract its value by knowing that the
slope obtained from the fit is just $-(\eps + 3)/\gamma$. By taking
the mean value of $\gamma$ from Table~\ref{tab:xirho}, 
we obtained $\epsrho = 1.15 \pm 0.03$ for (1.0,0.0), $\epsrho = 
0.48 \pm 0.08$ (0.2,0.0), and $\epsrho = 0.40 \pm 0.22$ for
(0.2,0.8). The low value of \epsrho\ for (0.2,0.8) mostly reflects
the fact that $\gamma$ is rather low at all epochs, in particular the
earliest one; for instance, $\epsrho = 0.96 \pm 0.20$ is obtained when
one does not consider the earliest epoch, which seems to be
a `transient' epoch.

A second way to compute \eps\ is assuming that in fact we do have
a two-point correlation function whose shape is independent of epoch
and its functionality with $1+z$ is a power-law. 
The \eps\ values we get when the evolution
is measured at 1 Mpc, \ie\ assuming $\xirho(1 \mpc,z) \propto 
(1+z)^{-(3+\eps)}$, are: 
$\epsrho = 1.04 \pm 0.09$ for (1.0,0.0),
$\epsrho = 0.18 \pm 0.12$ for (0.2,0.0), and
$\epsrho = 0.75 \pm 0.11$ for (0.2,0.8) which agree with
the above results within 1$\sigma$ for (1.0,0.0) and within
2$\sigma$ for (0.2,0.0) and (0.2,0.8).

We have also measured the evolution of \xirho\ at 200 \kpc. If
our scenarios are approximately self-similar we expect to measure the same 
rate of growth no matter which scale we use to measure it; 
i.e., we expect to get the same value of \eps.
That seems to apply to (1.0,0.0) model where an \eps\ value of 
$\epsrho = 1.03 \pm 0.14$ is computed when evolution is measured
at 0.2 \mpc. Notice, however, that $\epsrho = 0.73 \pm 0.11$ for 
(0.2,0.0) and  $\epsrho = 1.21 \pm 0.17$ for (0.2,0.8). 
This should not be surprising since these two
scenarios are far from satisfying a simple scaling
solution. 

So far, we have not mentioned what to expect for the evolution of 
\xirho, measured by the parameter \eps, from
the theoretical point of view. If the clustering is fixed
in comoving coordinates, \ie\ $\xi(x,a)= (r_0/x)^{\gamma}$, where
$r_0$ is the correlation length at the present time and $r=xa$, then
$\eps = \gamma -3$. On the other hand, in a highly non-linear regime,
$\xi \simgreat 200$, where bound gravitational units keep a fixed 
physical size, the clustering growth is the result of the
increasingly diluted background and $\eps = 0$. In the linear regime,
$\xi$ grows as the square of the growing mode of the 
linear density perturbations, $b$, 
which in the \omeone\ case it is proportional to the expansion factor, 
then $\xi(x,a) \propto a^2 x^{-\gamma}$ and therefore $\eps = \gamma -1$.
The \eps\ values that we compute from the evolution
of the correlation length or from the evolution of \xirho\ at
1 Mpc are slightly higher than those ones
expected from linear growing (for $b \neq a$ the \eps\ value
from linear growth is approximately the one obtained for \omeone\
multiplied by $[b(t_0)/a(t_0)]^2$, where $t_0$ is
the present time): $\eps = 0.94$ for (1.0,0.0), $\eps = 0.11$
for (0.2,0.0), and $\eps = 0.7$ for (0.2,0.8).

The evolution of \xirho\ has been measured from $z_e \sim 5$ to 
$z = 0.0$ and the \eps\ value would not have changed had we measured it
with a different $z_e$ value if both self-similarity and equation
(3) applied.
The \epsrho\ values, along with their error bars, as a function
of $z_e$ are shown in Figure~\ref{fig:epsrho}. 
Our previous results are the particular
case of $z_e \sim 5$. What is clear
from this figure is that the parameterization given by
equation (3) is a very rough approximation of $\xi(r,z)$.
This is not surprising at all since we are measuring the rate
of growth in different regimes; for example, at 
1 Mpc, the \epsrho\ value increases as $z_e$ decreases, because
evolution is passing from the linear to non-linear regime, whereas
at 0.2 Mpc the \epsrho\ value decreases because the rate of growth
is being measured almost only in the highly non-linear regime where 
we expect $\eps = 0$. The higher $\xi_{\rho \rho}(0.2 \mpc,0)$ 
is, the closer
one should expect the model be to the stable clustering regime. 
This agrees with
what we find for $\eps$ for the three models. Therefore, $\eps$ 
at 0.2 Mpc will be sensitive to parameters such as: $h$, $\sigma_8$
(normalization of $P(k)$), $\Gamma$. On the other hand, $\eps$ at
1 Mpc will depend much on the linear growth of density perturbations
which in turn is a function of \ome\ and \lam. It is expected to
be much less sensitive to $h$ and $\sigma_8$. 

\section{Evolution of \xihh}

Halos were initially identified using the friends of friends algorithm.
Any other group finding algorithm that suffers from the {\it overmerging}
problem is expected to give similar results. To show this we also used for
the (0.2,0.8) model the spherical overdensity algorithm, SO, by Lacey \&
Cole (1994) to identify halos. The mass
of each particle is given by $m_i = \ome \rho_c [\frac{L_{BOX}^3}{N}]$, where 
$\rho_c = 2.754 \times 10^{11}\ h^2 \msun / \mpc^3$, $L_{BOX}$ is the size of the
box in \mpc\ and $N = 128^3$ is the number of particles. Then, for (1.0,0.0)
$m_i$ is $1.31 \times 10^{11} \msun$, while for (0.2,0.0) and (0.2,0.8)
they are $2.62 \times 10^{10} \msun$ and $1.28 \times 10^{10} \msun$,
respectively. The galaxy-like
mass range used for halos was $5 m_i \le m_{halo}(\msun) \simless 10^{12}$. 
Because the mass of 
a particle in a low-density universe scenario is lower than in the
\omeone\ one, the number of particles contained in a halo of
a given mass in the former scenario is higher by 
a factor of $(h^2 \ome)^{-1}$. The link length we
used was $l=0.1$ ($l$ is in units of the mean 
interparticle spacing), which gives a minimun mean overdensity of
$\delta_{min} = 2000$, for the three models at each epoch. A mean
spherical overdensity of 2000 was chosen for the SO algorithm. Halos of
a rather high mean overdensity were chosen because they presumably suffer
less merging than their low mean overdensity counterparts. On the
other hand, a still higher
mean overdensity would produce a rather small number of halos, specially
at early epochs, and would make statistics very uncertain. 

The overdensity, $\delta_1$, reached
at the time of virialization of a spherical collapse is constant
for the (1.0,0.0) universe, 178, and it is a function of $t_{coll}$ for 
$\ome \neq 1$ universe, where $t_{coll}$ is the time where the 
collapse occurs; for example, at $z_{coll} \sim 5.0,$
$\delta_1 \sim 246$ for
(0.2,0.0) (it increases with time). 
An extra set of halos with a variable link length 
(increasing as we go to earlier epochs) was built for the (0.2,0.0) model.
The idea was to pick up halos whose difference 
$\delta_{min}-\delta_1$ at each epoch were the same for both (1.0,0.0) and (0.2,0.0).
No significant differences were found.

The evolution of \xirho\ and \xihh\ for the three models
is plotted in comoving coordinates in Figure~\ref{fig:xih1},
Figure~\ref{fig:xih2}, and Figure~\ref{fig:xih3}. 
In Figure~\ref{fig:xih3} we have
also plotted the evolution of \xihh\ as obtained by the
SO algorithm. The two earliest epochs were drawn out because
small numbers of halos make \xihh\ very uncertain. As far as
the rate of growth is concerned no significant differences can
be seen between the \xihh\ computed by using fof and the one
obtained with SO, hereafter results are given by using fof as
the group finding algorithm.
The function \xihh(x,z;\ome,\lam) is shaped by the interplay of four 
phenomena: (a) dynamical clustering, the amplitude
of \xihh\ increases due to gravitational clustering,
(b) merging, 
halos formed at high redshift may merge with other objects, increasing 
their mass out of the selected range, (c) formation, new halos in the
selected range are formed by accretion or merging, and 
(d) disruption, some halos are destroyed by tidal forces or
two-body encounters. Several
characteristics can be detected in Figure~\ref{fig:xih1}, Figure~\ref{fig:xih2},
and  Figure~\ref{fig:xih3}: 
(1) \xihh\ initially traces the filamentary structure of the universe;
i.e., halos are born ``naturally'' highly clustered,
producing a biased \xihh.
(2) The clustering of halos does not grow continuously, contrary to
the clustering of the mass density field. 
(3) If a lower mean overdensity were used we would have a 
greater overall suppression of \xihh.
This would reflect the fact that low-overdensity regions are more subject to 
merger and accretion, which decrease \xihh\ by moving halos out of the
selected range (\cite{carl91}). 
(4) The evolution
of \xihh\ is essentially fixed in comoving coordinates.
This is specially true for the models of low-density.
(5) Dynamical clustering does seem to drive
the clustering of halos at later epochs at scales $x \simgreat 0.7 \mpc$,
specially in the (1.0,0.0) model.

\subsection{Evolution of \xihh\ and the Parameter \eps}

A Table~\ref{tab:xihh} similar to Table~\ref{tab:xirho} 
was built for \xihh.
In this case the slope of \xihh\ was determined
using the separation range $1.0 \le x (\mpc) \le 10$. Unlike 
Table~\ref{tab:xirho}, in these tables we have preferred to show a 
mean \xihha. The 1$\sigma$ 
error bars are computed as follows: a set of 10 
realizations is built at each epoch and for each model, each one
being a random subset of its corresponding halo population. The number of
halos in each subset is one third of its
corresponding halo set. A mean \xihha\ with its typical deviation 
is then computed.
As was shown in Figure~\ref{fig:xih1} and Figure~\ref{fig:xih2} the comoving
correlation length at first decreases,
reaches a minimum and then starts increasing. This behavior does not
seem to depend on \ome\ (or on the mean overdensity of the halos).

In Figure~\ref{fig:xi0.1} we have plotted the evolution of 
\xihha\ (open squares) at 1 Mpc and 0.2 Mpc. The reason why we are choosing
200 \kpc\ is that it is the smallest scale where we could still measure
\xihh\ with a certain degree of confidence, \ie\ not being subject to
statistical noise produced by having too few particles. 
The evolution of \xirho\ is marked by asterisks. 

In Figure~\ref{fig:epshh} we have plotted \epshh\ as a function of
$z_e$. This Figure is the corresponding to Figure~\ref{fig:epsrho} but
for halos. 
It is interesting to see that the evolution at 1 Mpc from $z=0$ to $z_e \sim 1.3$
coincides with that of the mass density field for (1.0,0.0) model,
in agreement with the idea that in this redshift range halo clustering 
is essentially driven by the dynamical clustering of the mass density field. 
This is not sustained at small scales where mergers make the rate of growth
of \xihh\ lie below the rate of growth of \xirho. 
This was not discussed by BV because they
did not measure \xihh\ at these small scales. 
In the low-density models merging still appear to play a big
role in shaping the evolution of \xihh. 
At scales of 200 \kpc\ the growth of \xihh\ is
almost fixed in comoving coordinates for both scenarios, 
\ie\ $\eps \sim\ \gamma - 3$: for (1.0,0.0) $\gamma \sim 2.2$ and $\epshh 
\sim -0.8$, for (0.2,0.0) $\gamma \sim 1.6$ and $\epshh \sim -1.4$, and
for (0.2,0.8) $\gamma \sim 1.5$ and $\epshh \simless -1.5$. 
It is clear from what has been discussed so far 
that evolution in the two-point correlation can be a 
clear discrimation between models of low and high \ome. However,
little can be said between models with and without \lam\, unless
the evolution of \xigg\ is closer to that of the density field.

\section{Evolution of \sigrho\ and \sighh}

The 1--D pairwise velocity dispersion is defined to be
\begin{equation}
\sigma_1^2 = <({v_{\parallel}}-\bar{v_{\parallel}})^2>,
\end{equation}
where $v_{\parallel}$ stands for the component along the line
connecting the pair of the relative vector velocity. Quantities under
a bar or in brackets denotes mean. 
We have measured $\sigrho(1 \mpc,z)$ and computed a value for \eps\,
assuming that $\sigrho \propto (1+z)^{-\eps/2}$. The values for \eps\
we get are: $\eps_v = 1.78 \pm 0.13$ for (1.0,0.0), $\eps_v = 1.40
\pm 0.28$ for (0.2,0.0), and  $\eps_v = 2.72 \pm 0.28$ for (0.2,0.8)
(evolution is measured from $z \sim 5$ to $z = 0.0$). 
They are higher than the $\eps$ values we get
from the evolution of the correlation function. Our simulations
find more dynamical evolution than that predicted by the the CVT under
the assumption that the parameters $Q$ and $\gamma$ do not change with
epoch. 

It is interesting to see that our values for the present 1--D velocity
dispersion ($\simless 1000 \ \kms$ at 1 Mpc) for \ometwo\ are
higher than some values previously calculated in  
the literature (\eg\ \cite{defw}; \cite{kw}).
This, we believe, is due to three things: (1) the higher resolution of
our simulations, (2) the \sigei\ effect mentioned earlier (see
section \S 2), and (3) our rather high normalization. High
values for $\sigrho$ at small scales have been previously measured
(\eg\ Martel 1991). The value for the velocity dispersion obtained by our
open model is more than three times the observed galaxy velocity
dispersion.  

Table~\ref{tab:vel1} and Table~\ref{tab:vel2} show the evolution of the first and
second moment of the mass and halo velocity field for 
our models. We have preferred to show mean quantities for the halo population,
measured as we did for the two-point correlation function.
As expected, less evolution is  measured in \sighh\ as compared with 
\sigrho because halos are subject to merging and accretion.
On the other hand,  the picture we get from the
biased galaxy formation scenario, where
there is a time when galaxy formation ceases (or slows down) 
followed by dynamical clustering evolution, appears to be
just part of the story. It is here precisely where one expects halo dynamics
to trace better galaxy dynamics, and this view is supported by
the evolution of $\xihh$. The pairwise velocity
dispersion, however, decreases from $z= 0.5$ to $z = 0.0$
in all models. This may be due to
the high values of infall velocities encountered at $z = 0.5$.

\section{Summary}

We have measured the evolution of the two-point correlation function
and the pairwise velocity dispersion of the mass density field
and halo population. The evolution is parameterized mostly by
the \eps\ parameter. Our \eps\ values depend on the scale 
and the time period where evolution is measured, and for halos, they
also depend on their specified mean overdensity. Results were quoted
just for a mean overdensity of about 2000 because we believe halos
of this overdensity suffer less from the overmerging problem.
The \eps\ values for \xirho\ range from 0.4 (for (\ome,\lam)=(0.2,0.0)), 
when evolution is
measured at 0.2 Mpc, to 1.5, when evolution is measured at 1 Mpc 
(for (\ome,\lam)=(1.0,0.0),
both covering a period of time from $z \sim 1$ to present epoch.
The range of \epshh\ values covered by halos
is:  $-1.0 \simless \epshh \simless 1.5$ for (1.0,0.0),
$-1.4 \simless \epshh \simless -0.3$  for (0.2,0.0), and
$-2.1 \simless \epshh \simless -0.6$ for (0.2,0.8). 
The degree to which these results constrain the mean
density of the Universe depends on how well the evolution
of the galaxy clustering is traced by the evolution of
the mass density field or halo population. More and better observations
of $\xigg$ at diferent redshifts along with better numerical
determinations of its evolution are needed to constrain more 
the cosmological parameter space.

The correlation length, $r_0(z)$, was computed in two ways:
(1) using the ``standard'' definition $\xi(r_0,z) = 1$ and
(2) fitting $\xi$ to a power-law function at each epoch; \ie, 
assuming that $\xi(r,z) = (r/r_0(z))^{-\gamma}$, the shape
of $\xi$ is a power-law and does not change with time. 
The range of data used for the fit was from the 
softening length to the correlation length (as found by the first method),
to correspond to observations. We like to point out 
that the analytic evolution does not really do a good job
of predicting the details of the nonlinear correlation
function, errors as large as 50\% are obtained, so these
simple fits are considerable valuable.

The evolution of $\sigma(1 \mpc)$ was assumed to be a power-law with an
exponent $-\eps_v / 2$ (justified by the Cosmic Virial Theorem, CVT). 
All values found for $\eps_v$ (halo and mass density field) are
systematically higher than those predicted by the CVT; \ie we see more
evolution in the velocities than that predicted by the evolution of
$\xi$ and the CVT for constant $Q$ and $\gamma$.  

\acknowledgments

PC acknowledges a fellowship from Direcci\'on General
del Personal Acad\'emico, DGPA, UNAM, support from NSFRC at the University 
of Toronto, and a grant from DGPA through the project IN-109896. 
RGC and HMPC were supported by NSERC.
We thank Piotr Zembrowski for his help in
setting up the numerical computations. Its a pleasure to thank
Shaun Cole and Cedric Lacey for providing us with a copy of
their spherical overdensity algorithm. We are grateful to the anonymous
referee for his/her valuable comments.

\begin{deluxetable}{rrrrrrrrr}
\tablecolumns{9}
\tablenum{1}
\tablecaption{Evolutionary Parameters of \xirho\ \label{tab:xirho}}
\tablehead{\colhead{$z$} &  \colhead{$x_0$} &  
\colhead{$r_0$}\tablenotemark{a} & \colhead{$r_0$}\tablenotemark{b} 
& \colhead{$\sigma_{r_0}$} &
\colhead{$\gamma$} & \colhead{$\sigma_\gamma$} 
& \colhead{\xirho\ (1 \mpc)} & \colhead{\xirho\ (0.2 \mpc)} \\
\tablevspace{1.5ex} \cline{1-9} \tablevspace{1.5ex} 
\multicolumn{9}{c}{(\ome,\lam)=(1.0,0.0)}}
\startdata
0.00 & 4.63 & 4.63 & 4.86 & 0.14 & 1.86 & 0.05 & 35.958\phm{100} & 645.54\phm{100} \nl
0.52 & 3.17 & 2.09 & 2.20 & 0.06 & 1.94 & 0.06 &  5.490\phm{100} & 163.33\phm{100} \nl
0.91 & 2.62 & 1.37 & 1.42 & 0.03 & 1.98 & 0.06 &  1.933\phm{100} &  71.51\phm{100} \nl
1.48 & 2.12 & 0.86 & 0.85 & 0.02 & 2.02 & 0.05 &  0.745\phm{100} &  23.84\phm{100} \nl
2.85 & 1.37 & 0.36 & 0.35 & 0.01 & 2.12 & 0.03 &  0.150\phm{100} &   2.85\phm{100} \nl
4.87 & 0.85 & 0.15 & 0.14 & 0.00 & 1.74 & 0.07 &  0.025\phm{100} &   0.64\phm{100} \nl
\cutinhead{(\ome,\lam)=(0.2,0.0)}
0.00 & 6.05 & 6.07 & 6.06 & 0.12 & 1.91 & 0.04 & 45.48\phm{100} & 1000.1\phm{100} \nl
0.50 & 4.99 & 3.32 & 3.37 & 0.07 & 1.91 & 0.04 &  9.72\phm{100} &  263.8\phm{100} \nl
0.99 & 4.06 & 2.04 & 2.16 & 0.05 & 1.91 & 0.05 &  3.46\phm{100} &  110.9\phm{100} \nl
1.40 & 3.68 & 1.54 & 1.59 & 0.03 & 1.88 & 0.03 &  2.00\phm{100} &   52.3\phm{100} \nl
2.74 & 2.47 & 0.66 & 0.70 & 0.02 & 1.88 & 0.03 &  0.56\phm{100} &    8.2\phm{100} \nl
5.00 & 1.49 & 0.25 & 0.27 & 0.00 & 1.67 & 0.05 &  0.14\phm{100} &    1.4\phm{100} \nl
\cutinhead{(\ome,\lam)=(0.2,0.8)}
0.00 & 6.42 & 6.42 & 9.31 & 0.53 & 1.55 & 0.06 & 44.21\phm{100} & 453.6\phm{100} \nl
0.49 & 5.33 & 3.58 & 4.03 & 0.21 & 1.56 & 0.05 &  9.58\phm{100} &  128.0\phm{100} \nl
0.97 & 3.74 & 1.90 & 2.05 & 0.08 & 1.54 & 0.04 &  2.73\phm{100} &  43.0\phm{100} \nl
1.60 & 2.42 & 0.93 & 0.90 & 0.03 & 1.57 & 0.03 &  0.91\phm{100} &   11.6\phm{100} \nl
2.99 & 1.08 & 0.27 & 0.25 & 0.03 & 1.31 & 0.06 &  0.22\phm{100} &    1.4\phm{100} \nl
4.00 & 0.34 & 0.07 & 0.25 & 0.03 & 0.87 & 0.06 &  0.11\phm{100} &    0.7\phm{100} \nl
\enddata
\tablenotetext{a}{The correlation length, in physical units, computed by
solving the equation $\xi(r_0)=1.0$. The $\log \xi$ was simply
interpolated linearly between the radii $r'$ and $r''$, where $\xi(r')
< 1$ and $\xi(r'') \ge 1$. }
\tablenotetext{b}{ The correlation length, in physical units, computed by
fitting the $log \xi\ vs\ \log r$ through least squares.}
\end{deluxetable}

\begin{deluxetable}{cccccccrrrr}
\tablecolumns{11}
\tablenum{2}
\tablecaption{Evolutionary Parameters of \xihh\ \label{tab:xihh}}
\tablehead{\colhead{$z$} & \colhead{$x_0$} & \colhead{$r_0$} & 
\colhead{$r_0$} & \colhead{$\sigma_{r_0}$} &
\colhead{$\gamma$} & \colhead{$\sigma_\gamma$} & 
\colhead{$\xihha(1 \ \mpc)$} & \colhead{$\sigma_\xi$} & 
\colhead{$\xihha(0.2 \ \mpc)$} & \colhead{$\sigma_\xi$} \\ 
\tablevspace{1.5ex} \cline{1-11} \tablevspace{1.5ex} 
\multicolumn{11}{c}{(\ome,\lam)=(1.0,0.0)}}
\startdata
0.00 & 4.43 & 4.43 & 4.42 & 0.08 & 2.41 & 0.04 & 35.70\phm{100} & 1.34 & 188.8\phm{1000} &  23.4 \nl
0.52 & 3.03 & 2.00 & 2.10 & 0.07 & 2.25 & 0.06 &  5.53\phm{100} & 0.24 &  84.1\phm{1000} &  7.8 \nl
0.91 & 2.67 & 1.40 & 1.41 & 0.06 & 2.16 & 0.06 &  1.91\phm{100} & 0.09 &  47.5\phm{1000} &  2.9 \nl
1.48 & 2.44 & 0.99 & 0.93 & 0.04 & 2.17 & 0.07 &  0.93\phm{100} & 0.07 &  24.4\phm{1000} &  1.4 \nl
2.85 & 2.24 & 0.58 & 0.55 & 0.04 & 2.26 & 0.14 &  0.33\phm{100} & 0.04 &   6.6\phm{1000} &  0.6 \nl
4.87 & 2.48 & 0.42 & 0.40 & 0.04 & 2.29 & 0.14 &  0.15\phm{100} & 0.03 &   4.0\phm{1000} &  1.0 \nl
\cutinhead{(\ome,\lam)=(0.2,0.0)}
0.00 & 5.14 & 5.16 & 5.02 & 0.08 & 1.93 & 0.02 & 21.22\phm{100} & 0.79 & 113.3\phm{1000} & 9.7 \nl
0.50 & 4.41 & 2.94 & 2.99 & 0.07 & 1.79 & 0.03 &  7.44\phm{100} & 0.35 &  61.2\phm{1000} & 2.7 \nl
0.99 & 4.29 & 2.15 & 2.13 & 0.06 & 1.71 & 0.04 &  3.38\phm{100} & 0.15 &  43.7\phm{1000} & 3.5 \nl
1.40 & 4.14 & 1.73 & 1.72 & 0.04 & 1.66 & 0.03 &  2.25\phm{100} & 0.08 &  34.8\phm{1000} & 3.2 \nl
2.74 & 4.14 & 1.11 & 1.08 & 0.02 & 1.60 & 0.02 &  1.13\phm{100} & 0.06 &  17.3\phm{1000} & 1.4 \nl
5.00 & 4.45 & 0.74 & 0.73 & 0.01 & 1.60 & 0.02 &  0.61\phm{100} & 0.04 &   7.6\phm{1000} & 0.7 \nl
\cutinhead{(\ome,\lam)=(0.2,0.8)}
0.00 & 6.96 & 6.96 & 7.08 & 0.10 & 1.99 & 0.02 & 41.81\phm{100} & 2.59 & 225.9\phm{1000} & 26.3 \nl
0.49 & 7.54 & 5.06 & 4.73 & 0.20 & 1.86 & 0.06 & 19.87\phm{100} & 1.34 & 175.4\phm{1000} & 16.8 \nl
0.97 & 7.02 & 3.56 & 3.53 & 0.09 & 1.67 & 0.03 &  7.90\phm{100} & 0.27 & 120.7\phm{1000} &  8.5 \nl
1.60 & 6.79 & 2.61 & 2.54 & 0.06 & 1.49 & 0.03 &  3.74\phm{100} & 0.20 &  66.2\phm{1000} &  7.6 \nl
2.99 & 8.58 & 2.15 & 2.08 & 0.05 & 1.52 & 0.03 &  3.05\phm{100} & 0.26 &  46.2\phm{1000} & 11.9 \nl
4.00 & 11.87 & 2.38 & 2.21 & 0.14 & 1.55 & 0.08 & 3.32\phm{100} & 1.55 &  43.4\phm{1000} & 65.8 \nl
\enddata
\end{deluxetable}

\begin{deluxetable}{crrr}
\tablecolumns{4}
\tablewidth{8cm}
\tablenum{3}
\tablecaption{Evolution of $\bar{v_\parallel}$ and $\sigrho$
\tablenotemark{a} \label{tab:vel1}}
\tablehead{\colhead{$z$} &  \colhead{$H(z) r$} & \colhead{$\bar{v_\parallel}$} & \colhead{$\sigrho$} \\ 
\tablevspace{1.5ex} \cline{1-4} \tablevspace{1.5ex} 
\multicolumn{4}{c}{(\ome,\lam)=(1.0,0.0)}}
\startdata
0.00 &  100 & 157 & 1485 \nl
0.52 &  187 & 384 & 1192 \nl
0.91 &  263 & 450 &  903 \nl
1.48 &  390 & 354 &  688 \nl
2.85 &  756 & 172 &  467 \nl
4.87 & 1421 &  71 &  344 \nl
\cutinhead{(\ome,\lam)=(0.2,0.0)}
0.00 & 100 &  164 & 981 \nl
0.50 & 158 &  265 & 879 \nl
0.99 & 218 &  288 & 782 \nl
1.40 & 271 &  347 & 649 \nl
2.74 & 466 &  191 & 385 \nl
5.00 & 848 &  107 & 270 \nl
\cutinhead{(\ome,\lam)=(0.2,0.8)}
0.00 &  70 &  140 & 728 \nl
0.49 &  85 &  140 & 568 \nl
0.97 & 107 &  211 & 399 \nl
1.60 & 146 &  121 & 219 \nl
2.99 & 257 &   57 & 120 \nl
4.00 & 355 &   39 &  97 \nl
\enddata
\tablenotetext{a}{The velocities are measured at a physical constant
separation of 1 \mpc\ in physical units, \kms.}
\end{deluxetable}

\begin{deluxetable}{rrrrrr}
\tablecolumns{6}
\tablenum{4}
\tablewidth{11cm}
\tablecaption{Evolution of $<\bar{v_\parallel}>$ and $<\sighh>$ 
\label{tab:vel2}}
\tablehead{\colhead{$z$} & \colhead{$H(z)r$} & \colhead{$<\bar{v_\parallel}>$} & \colhead{$\sigma_{<\bar{v_\parallel}>}$} & \colhead{$<\sighh>$} 
& \colhead{$\sigma_{<\sighh>}$} \\ 
\tablevspace{1.5ex} \cline{1-6} \tablevspace{1.5ex} 
\multicolumn{6}{c}{(\ome,\lam)=(1.0,0.0)}}
\startdata
0.00 &  100 & 116 & 31 & 861 & 29 \nl
0.52 &  187 & 318 & 55 & 924 & 35 \nl
1.48 &  390 & 355 & 21 & 557 & 14 \nl
2.85 &  756 & 241 & 14 & 413 &  8 \nl
4.87 & 1421 & 176 & 14 & 354 & 11 \nl
\cutinhead{(\ome,\lam)=(0.2,0.0)}
0.00 &  100 & 110 & 10 & 460 & 10 \nl
0.52 &  158 & 243 & 17 & 504 & 14 \nl
1.48 &  271 & 268 & 12 & 425 & 16 \nl
2.85 &  466 & 226 &  9 & 335 &  9 \nl
4.87 &  848 & 196 &  5 & 278 &  4 \nl
\cutinhead{(\ome,\lam)=(0.2,0.8)}
0.00 &  70 &  77 &  9 & 332 & 11 \nl
0.49 &  84 & 137 & 14 & 370 &  9 \nl
0.97 & 107 & 251 & 13 & 341 & 10 \nl
1.60 & 146 & 162 & 10 & 200 &  8 \nl
2.99 & 257 & 156 &  9 & 149 &  9 \nl
4.00 & 355 & 142 & 20 & 127 & 11 \nl 
\enddata
\tablecomments{As in Table 5a the velocities here are also measured 
at 1 \mpc}
\end{deluxetable}

\newpage

\figcaption[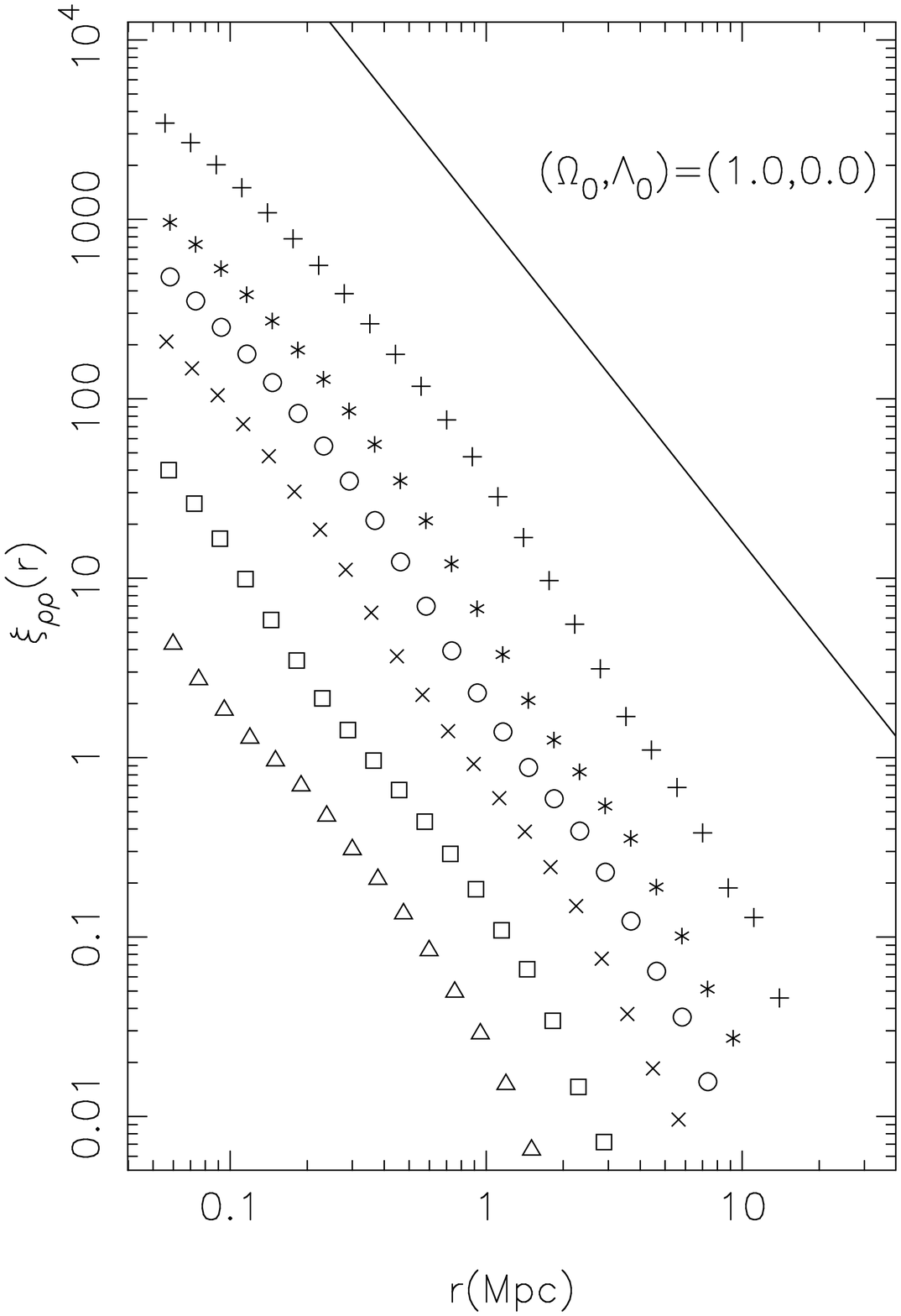]{The evolution of the two-point correlation function 
of the density field in physical coordinates measured at different epochs, 
from $z = 0.0$ to $z \sim 5.0$ (from top to bottom), for the \omeone\ 
scenario. The line is a power-law with exponent -1.8. \label{fig:xirho1}}

\figcaption[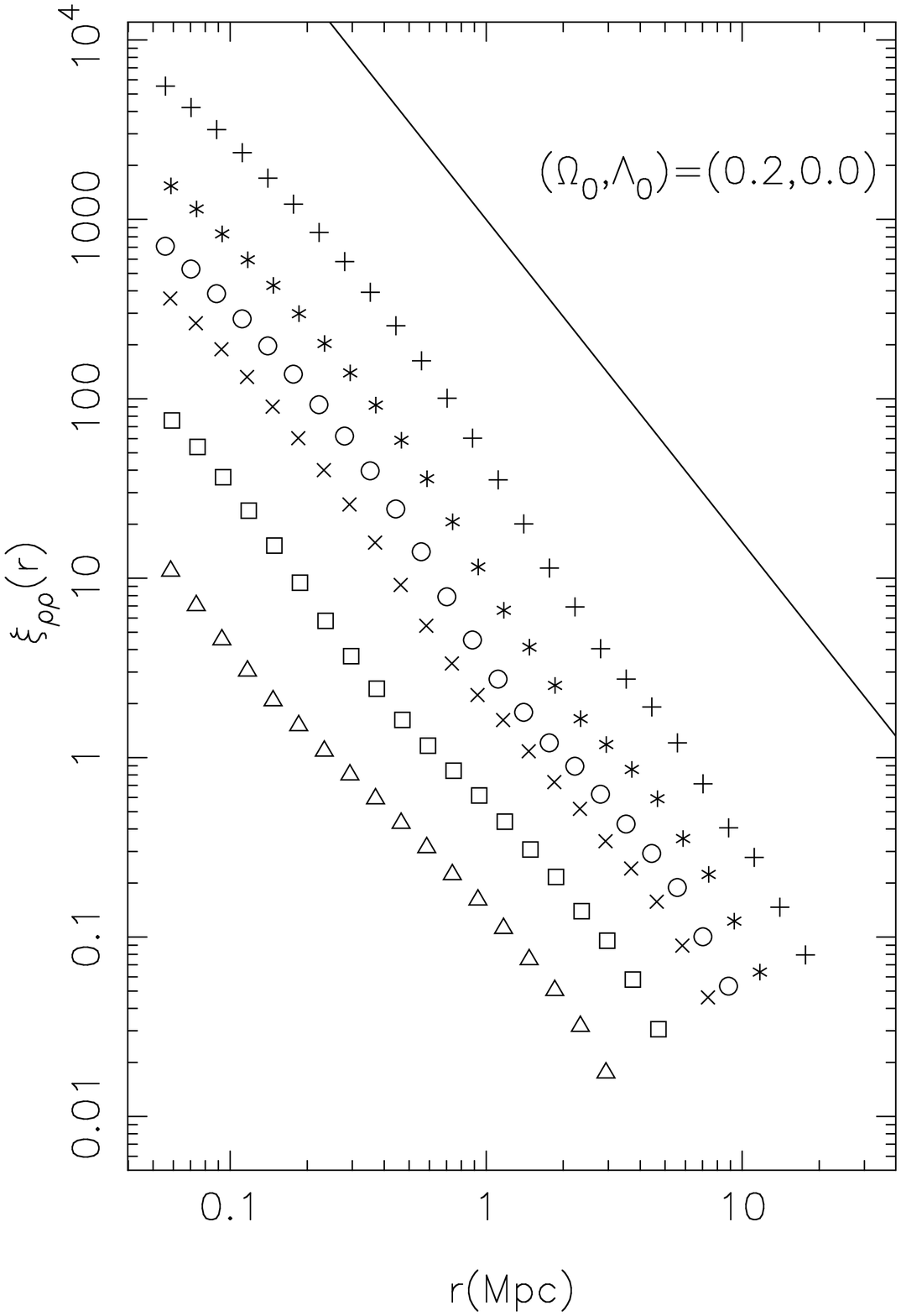]{The same as Figure 1 but for \ometwo. 
\label{fig:xirho2}}

\figcaption[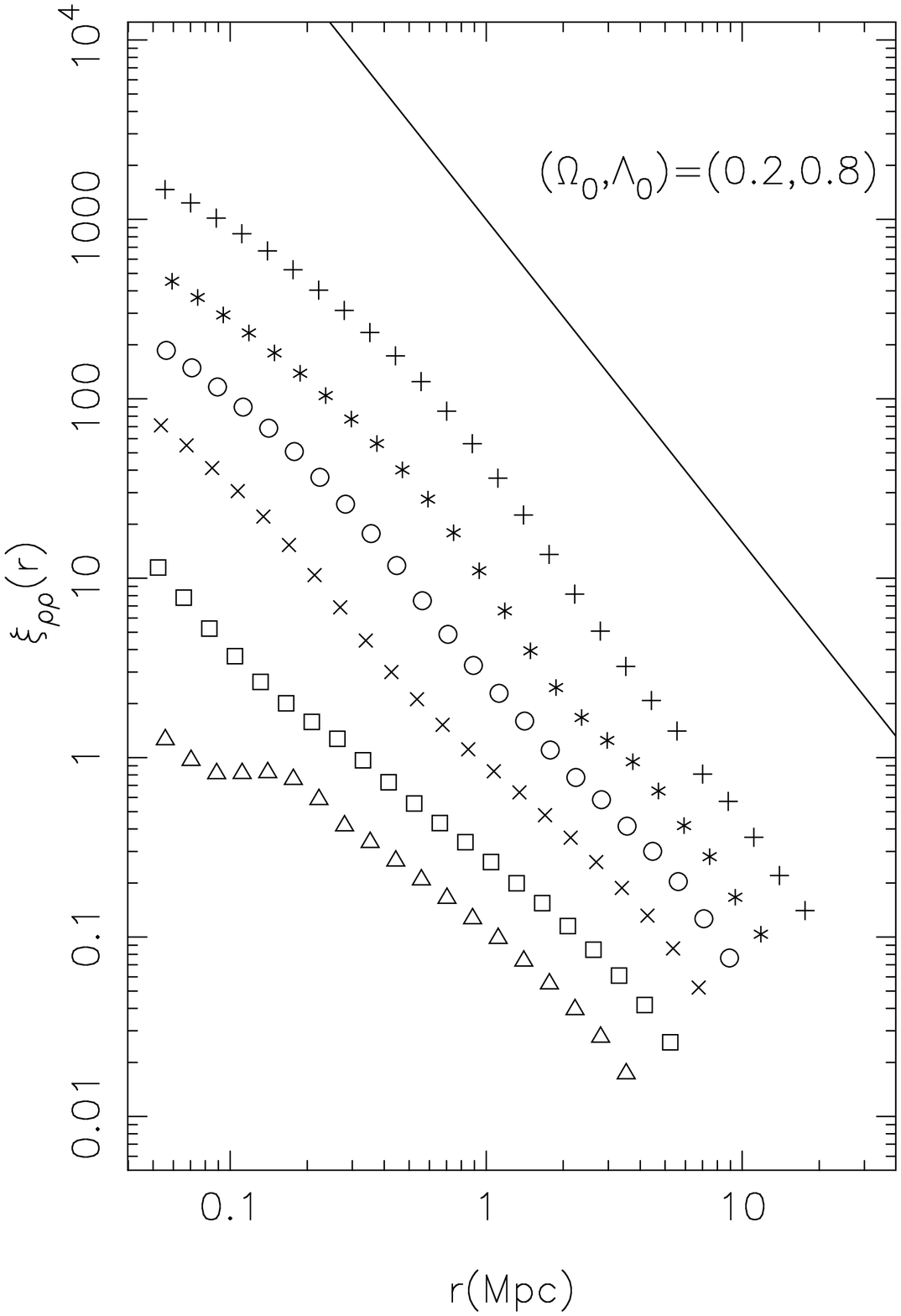]{The same as Figure 1 but for \ometwo. 
\label{fig:xirho3}}

\figcaption[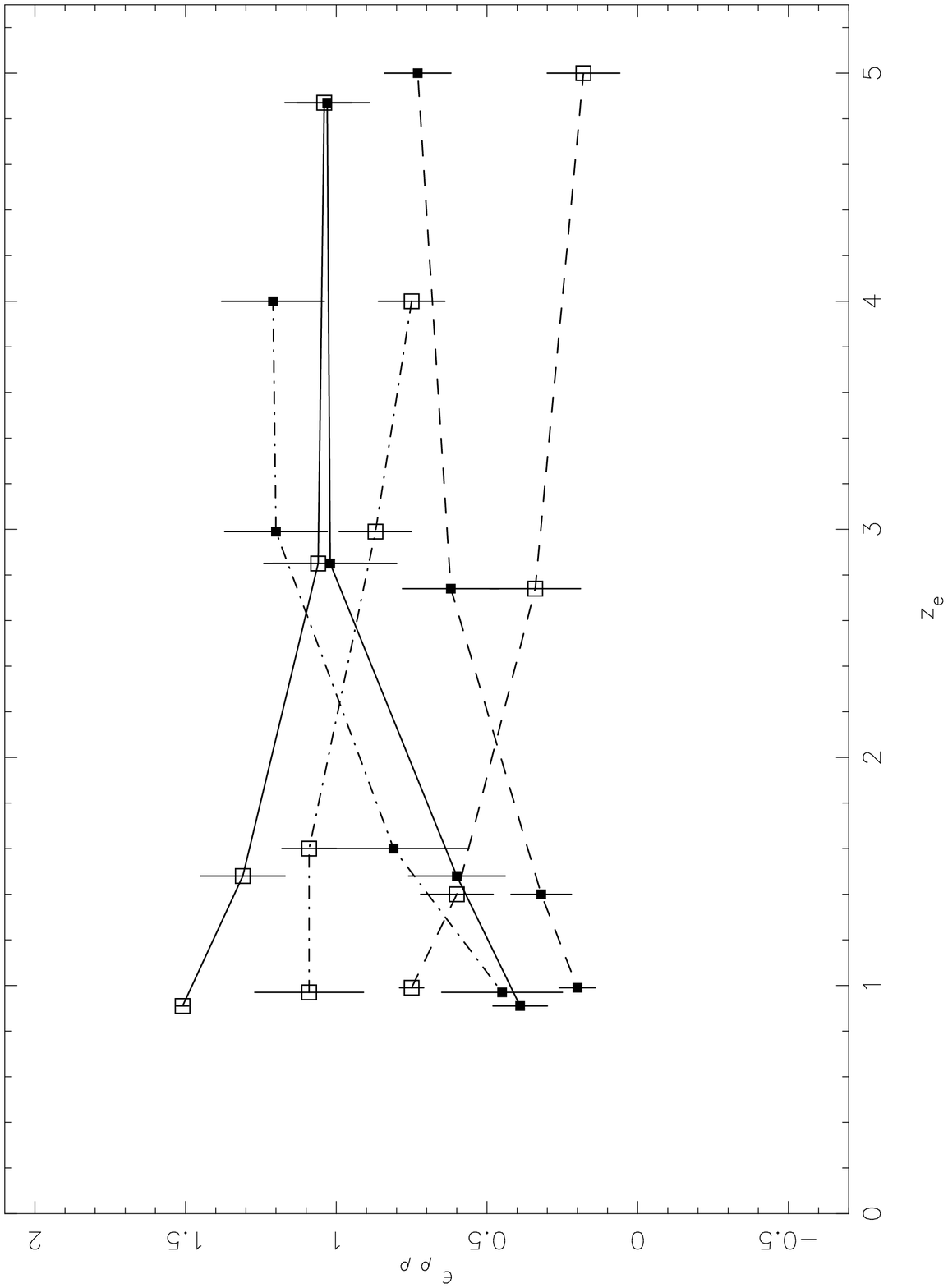]{ The evolution of \xirho\ from $z=z_e$ to $z=0$ 
measured by the \epsrho\ parameter is plotted as a function of $z_e$. Different
types of lines denote different models: solid line for (1.0,0.0), dot-dashed
line for (0.2,0.8), and dashed line for (0.2,0.0). Open and filled squares measure
evolution at 1 Mpc and 0.2 Mpc, respectively. \label{fig:epsrho}}

\figcaption[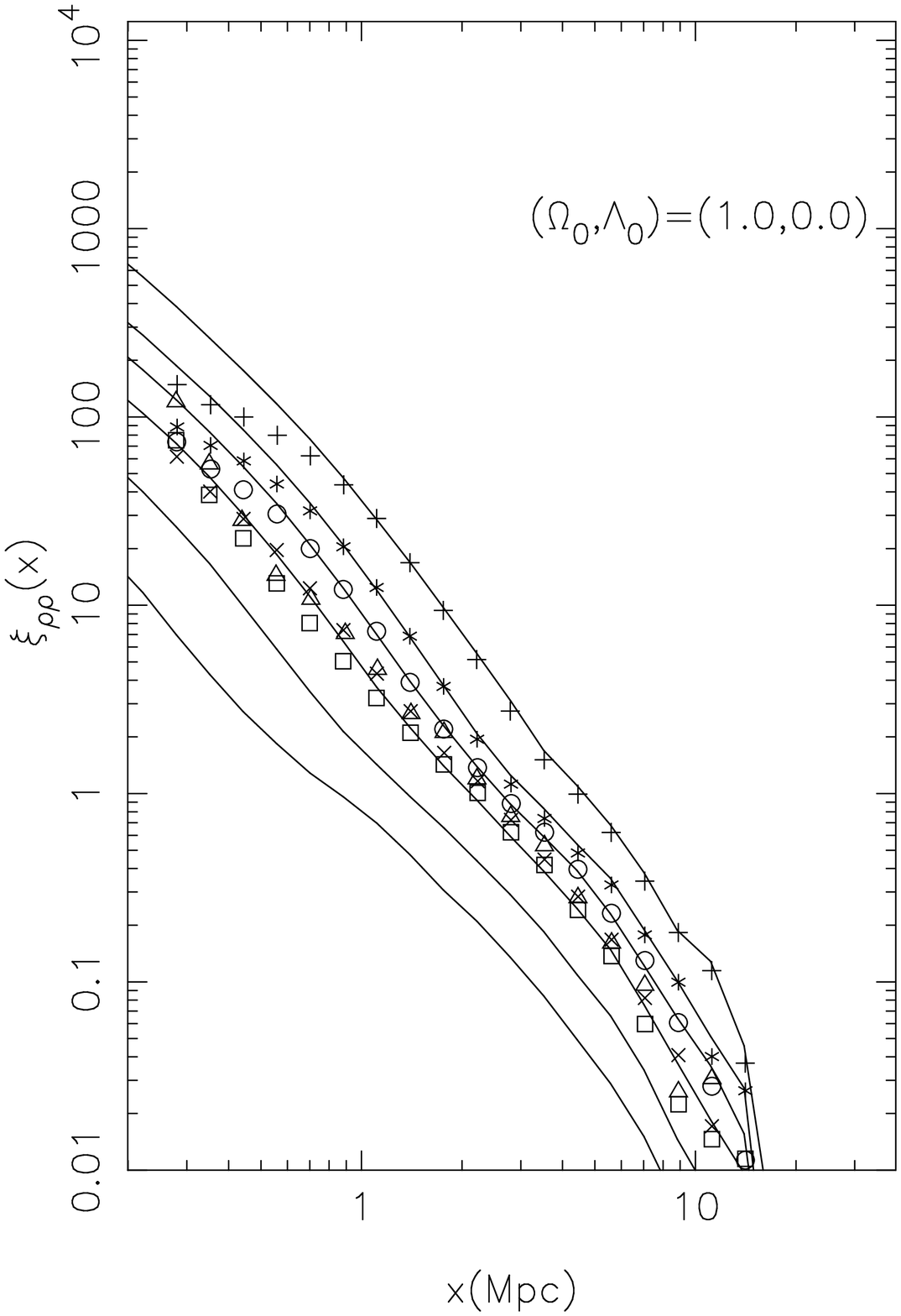]{The evolution of the two-point correlation function
of the mass density field (lines) and halo population (symbols) 
(\xirho\ and \xihh\, respectively) in comoving coordinates for 
(\ome,\lam)=(1.0,0.0). Epochs are marked as in Fig. 1. \label{fig:xih1}}

\figcaption[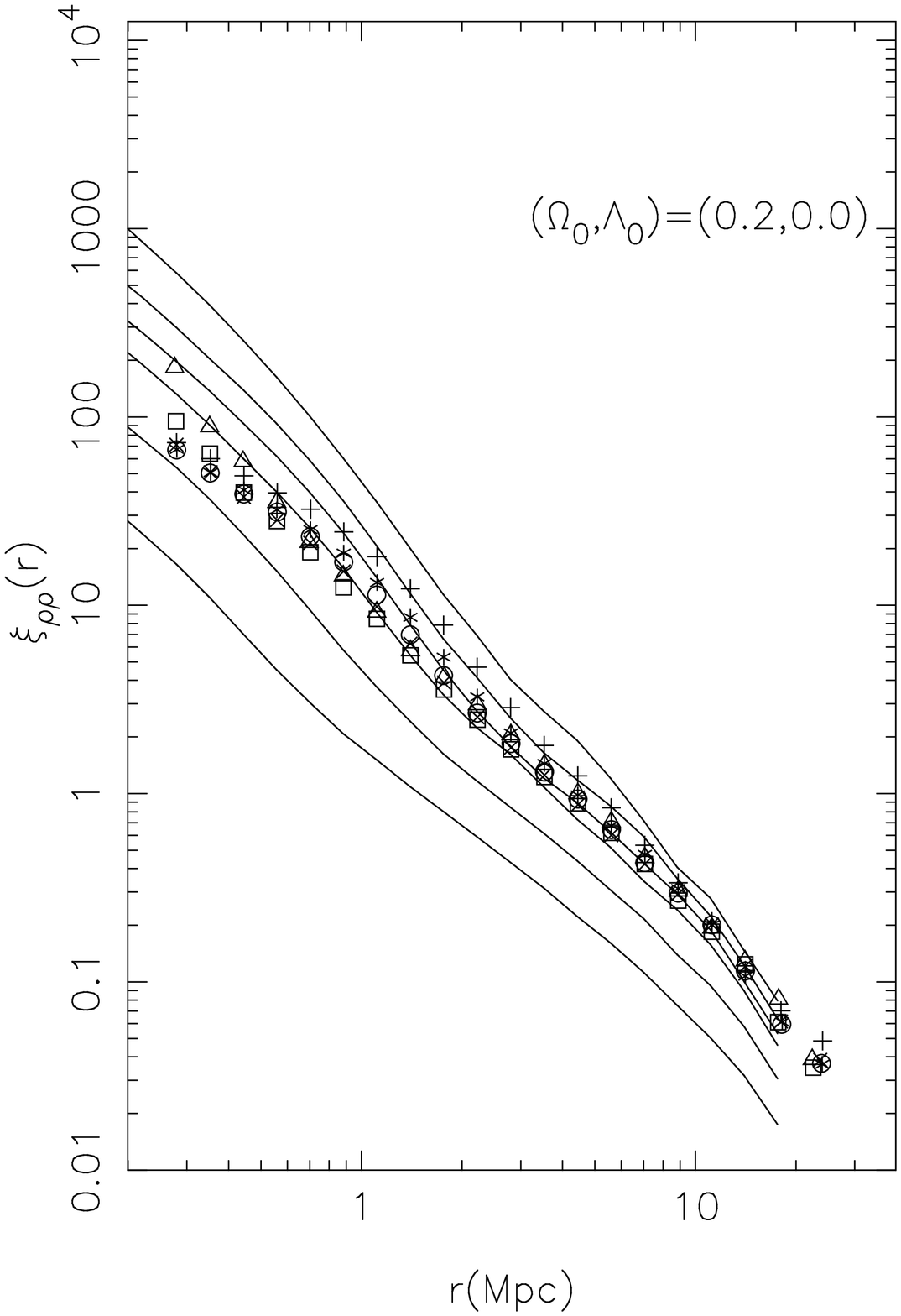]{same as Figure 4 for (\ome,\lam)=(0.2,0.0). 
\label{fig:xih2}}

\figcaption[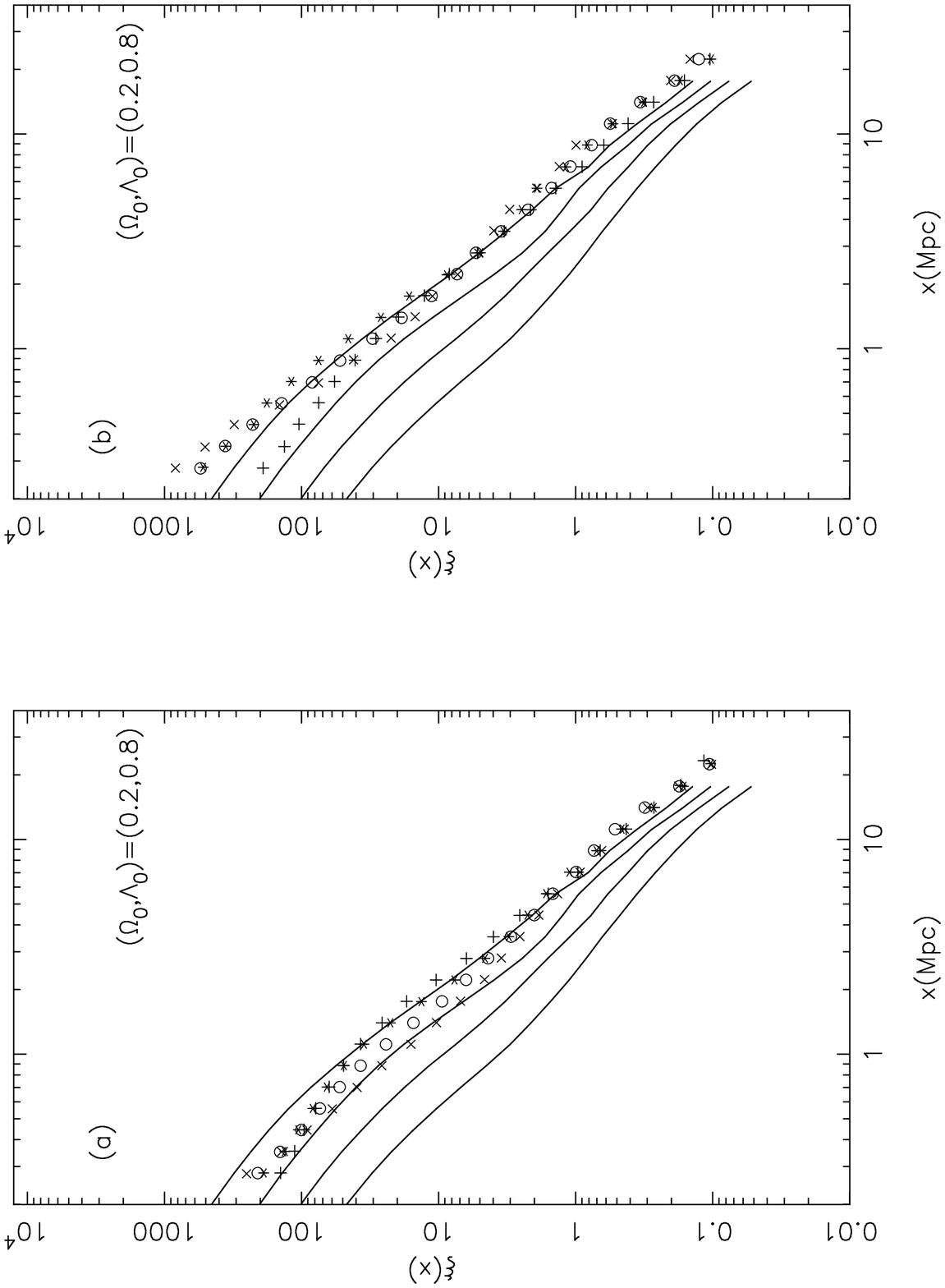]{same as Figure 4 and 5 for (\ome,\lam)=(0.2,0.8). 
In (a) \xihh\ was found by using {\it friends of friends} as the
group finding algorithm and (b) \xihh\ as found by the
spherical overdensity algorithm. \label{fig:xih3}}

\figcaption[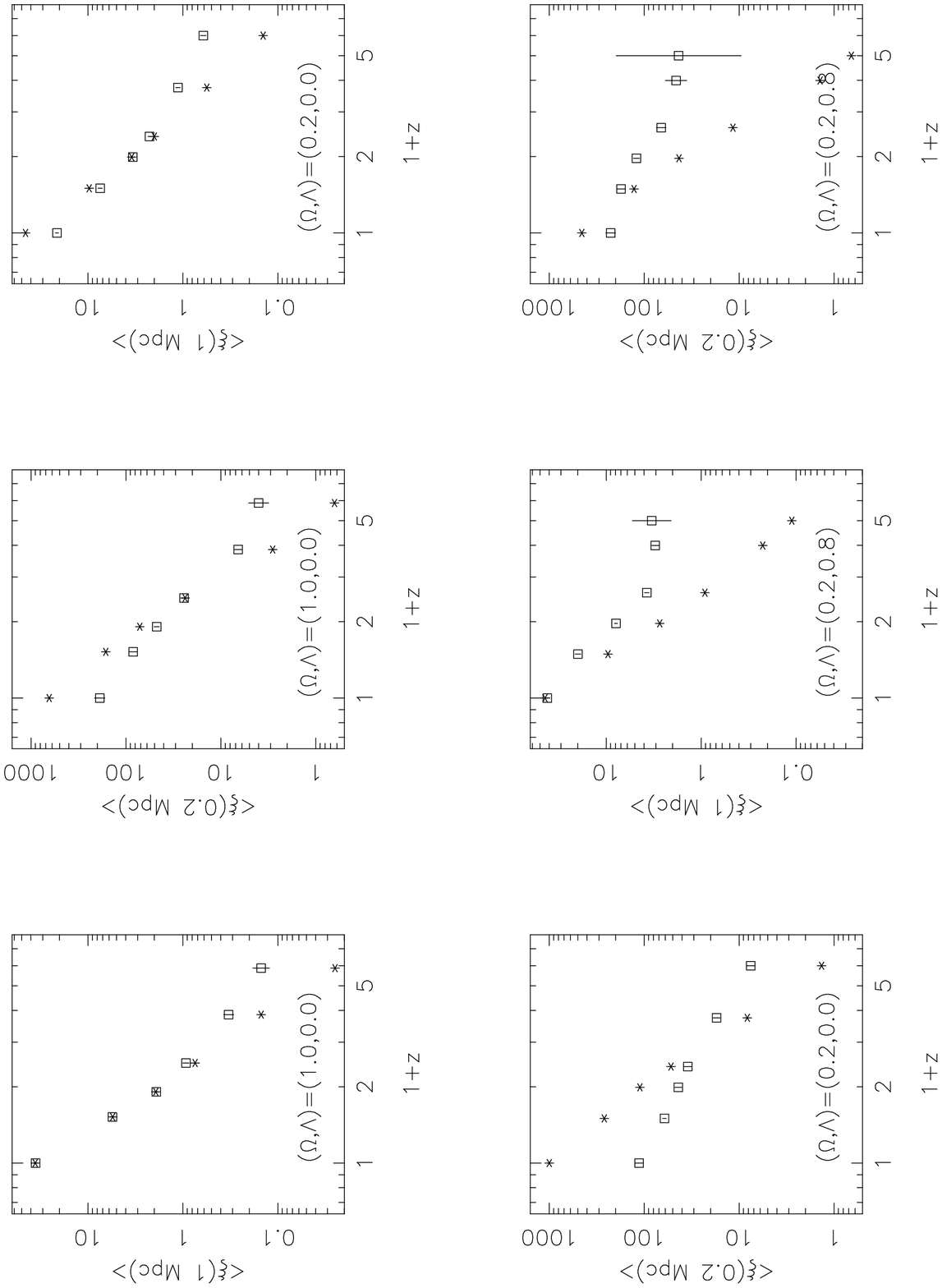]{Evolution of the two-point correlation function 
at 1 \mpc\ and 0.2 \mpc\ of the halo population (squares, $l=0.1$) and 
the mass density field (asterisks). \label{fig:xi0.1}}

\figcaption[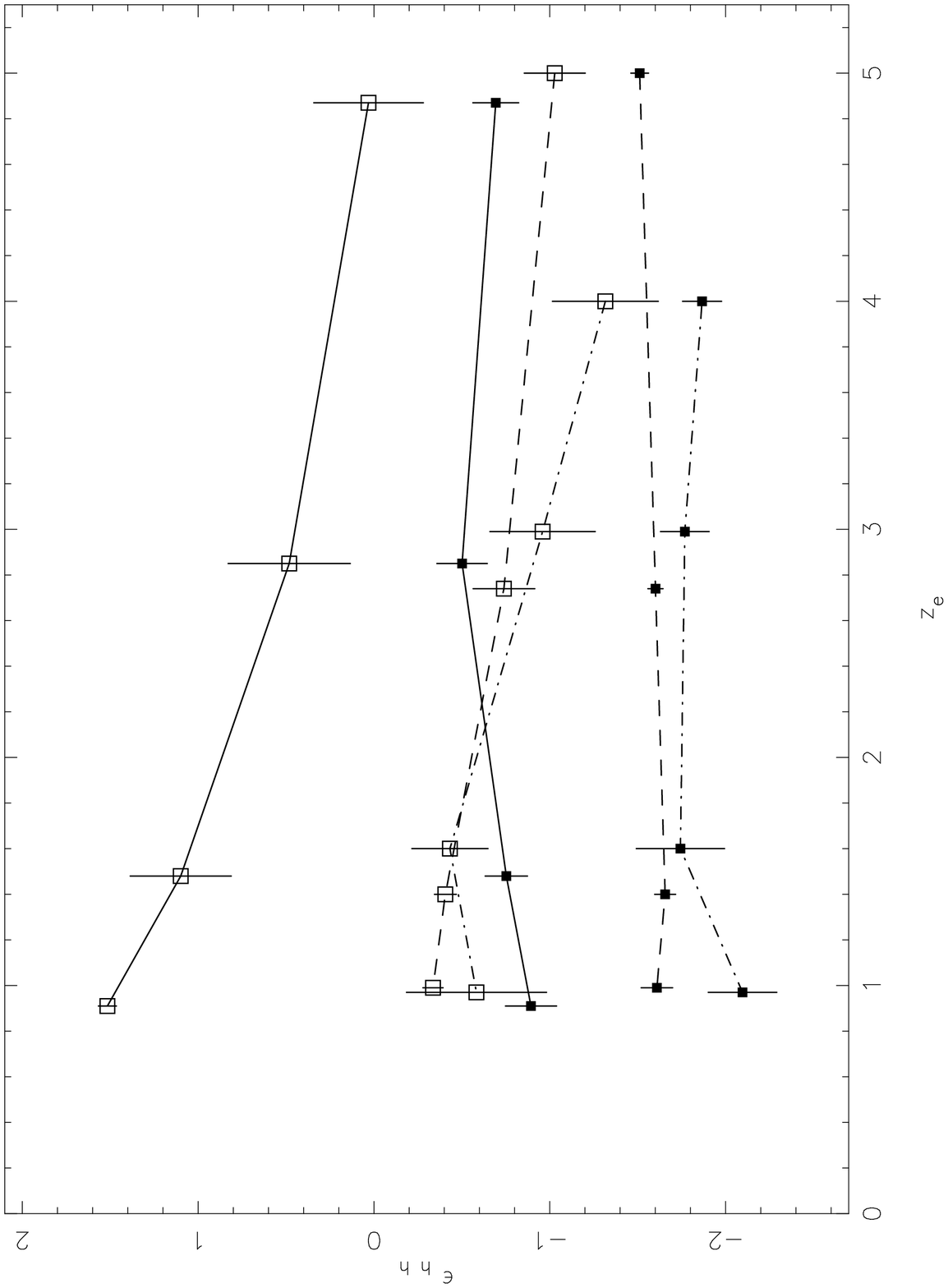]{Evolution of the halo two-point correlation function
as measured by \epshh. Symbols and lines are the same as in Figure 4.
\label{fig:epshh}}

\begin{figure}
\figurenum{\ref{fig:xirho1}}
\plotone{fig1.eps}
\end{figure}

\begin{figure}[h]
\figurenum{\ref{fig:xirho2}}
\plotone{fig2.eps}
\end{figure}

\begin{figure}[h]
\figurenum{\ref{fig:xirho3}}
\plotone{fig3.eps}
\end{figure}

\begin{figure}[h]
\figurenum{\ref{fig:epsrho}}
\plotone{fig4.eps}
\end{figure}

\begin{figure}[h]
\figurenum{\ref{fig:xih1}}
\plotone{fig5.eps}
\end{figure}

\begin{figure}[h]
\figurenum{\ref{fig:xih2}}
\plotone{fig6.eps}
\end{figure}

\begin{figure}[h]
\figurenum{\ref{fig:xih3}}
\plotone{fig7.eps}
\end{figure}
        

\begin{figure}[h]
\figurenum{\ref{fig:xi0.1}}
\plotone{fig8.eps}
\end{figure}

\begin{figure}[h]
\figurenum{\ref{fig:eps}}
\plotone{fig9.eps}
\end{figure}

\end{document}